\documentclass[aps,prb,twocolumn,superscriptaddress,showpacs,showkeys,floatfix]{revtex4}

\usepackage{graphicx,color}
\usepackage{multirow,slashbox}

\newcommand{\jwj}[1]{\textcolor{blue}{#1}}

\graphicspath{{figs/}}
\begin{document}
\title{Molecular Dynamics Simulations of Single-Layer Molybdenum Disulphide (MoS$_{2}$): Stillinger-Weber Parametrization, Mechanical Properties, and Thermal Conductivity}
\author{Jin-Wu Jiang}
    \altaffiliation{Electronic address: jinwu.jiang@uni-weimar.de}
    \affiliation{Institute of Structural Mechanics, Bauhaus-University Weimar, Marienstr. 15, D-99423 Weimar, Germany}
\author{Harold S. Park}
    \altaffiliation{Electronic address: parkhs@bu.edu}
    \affiliation{Department of Mechanical Engineering, Boston University, Boston, Massachusetts 02215, USA}
\author{Timon Rabczuk}
    \altaffiliation{Electronic address: timon.rabczuk@uni-weimar.de}
    \affiliation{Institute of Structural Mechanics, Bauhaus-University Weimar, Marienstr. 15, D-99423 Weimar, Germany}
    \affiliation{School of Civil, Environmental and Architectural Engineering, Korea University, Seoul, South Korea }

\date{\today}
\begin{abstract}

We present a parameterization of the Stillinger-Weber potential to describe the interatomic interactions within single-layer MoS$_{2}$ (SLMoS$_{2}$). The potential parameters are fitted to an experimentally-obtained phonon spectrum, and the resulting empirical potential provides a good description for the energy gap and the crossover in the phonon spectrum.  Using this potential, we perform classical molecular dynamics simulations to study chirality, size, and strain effects on the Young's modulus and the thermal conductivity of SLMoS$_{2}$. We demonstrate the importance of the free edges on the mechanical and thermal properties of SLMoS$_{2}$ nanoribbons.  Specifically, while edge effects are found to reduce the Young's modulus of SLMoS$_{2}$ nanoribbons, the free edges also reduce the thermal stability of SLMoS$_{2}$ nanoribbons, which may induce melting well below the bulk melt temperature.  Finally, uniaxial strain is found to efficiently manipulate the thermal conductivity of infinite, periodic SLMoS$_{2}$.

\end{abstract}

\maketitle
\pagebreak

\section{Introduction}

As the first two-dimensional material, the physical properties of graphene have been extensively investigated, as shown in the reviews in Refs.~\onlinecite{GeimAK2007nm,LiD2008sci,GeimAK2009sci,CastroNAH,RaoCNR,Balandin2011nm}. Inspired by the novel physical properties of two-dimensional graphene, there has been increasing interest in studying other similar two-dimensional and one-atom-thick layered materials,\cite{NovoselovKS2005pnas} particularly as these more recently studied two-dimensional materials may have superior properties to graphene. For instance, MoS$_{2}$ is a semiconductor with a bulk bandgap above 1.2~{eV},\cite{KamKK} which can be further manipulated by changing its thickness.\cite{MakKF} This finite bandgap is a key reason for the excitement surrounding MoS$_{2}$ as compared to graphene as graphene has a zero bandgap unless strain,\cite{PereiraVM} or other gap-opening engineering is performed.\cite{NovoselovKS2005nat}  Because of its direct bandgap and also its well-known properties as a lubricant, MoS$_{2}$ has attracted considerable attention in recent years.\cite{WangQH2012nn,ChhowallaM}

Similar to graphene, single-layer MoS$_{2}$ (SLMoS$_{2}$) can be obtained by the mechanical exfoliation technique. Recent studies have shown that the SLMoS$_{2}$ is a promising alternative to graphene in some electronic fields. Radisavljevic et al.\cite{RadisavljevicB2011nn} demonstrated the application of SLMoS$_{2}$ as a good transistor, which has a large mobility above 200~{cm$^{2}$V$^{-1}$s$^{-1}$}. This large mobility is realized through the usage of a hafnium oxide gate dielectric. Considering its large intrinsic direct bandgap of 1.8~{eV},\cite{MakKF} SLMoS$_{2}$ may also find application in optoelectronics, energy harvesting, and other nano-material fields.\cite{JoensenP,HelvegS,MakKF,LeeC,RadisavljevicB2011nn,PopovI,AtacaC,SanchezAM,SahooS,YinZ,ChangK,AtcaC,RadisavljevicB2012apl,Castellanos-GomezA,ColemanJN,LiuKK,LeD,BrivioJ}

Quite recently, the thermal transport properties of SLMoS$_{2}$ have started to attract attention, due to the superior thermal conductivity found in graphene.\cite{BalandinAA2008,NikaDL2009prb,NikaDL2009apl,Balandin2011nm,JiangJW2009direction} First-principles calculations were performed to investigate the thermal transport in the SLMoS$_{2}$ in the ballistic transport region.\cite{HuangW,JiangJW2013mos2} For diffusive thermal transport, a force-field based molecular dynamics (MD) simulation was performed by Varshney et al.\cite{VarshneyV}

For theoretical investigations of SLMoS$_{2}$, first-principles calculations are the most accurate approach, due to their ability to provide quantum mechanically-based predictions for various properties of SLMoS$_{2}$. However, it is well-known that first principles techniques cannot simulate more than around a few thousand atoms, which poses serious limitations for comparisons to experimental studies, which typically occur on the micrometer length scale.  For such larger sizes, classical molecular dynamics (MD) simulations are desirable.  However, the key ingredient for accurate MD simulations is an empirical interatomic potential from which the forces that are needed in Newton's equations of motion can be derived.  

Several such interatomic potentials have been proposed for MoS$_{2}$ and SLMoS$_{2}$. In 1975, Wakabayashi et al.\cite{WakabayashiN} developed a valence force field (VFF) model to calculate the phonon spectrum of the bulk MoS$_{2}$. This model has been applied to study the lattice dynamics properties of MoS$_{2}$ based materials.\cite{JimenezSS,DobardzicE,DamnjanovicM2008mmp} The VFF model is a harmonic potential, which is only valid for predictions of the elastic properties of MoS$_{2}$, and thus cannot be used in general for large-scale MD simulations of thermal transport or mechanical deformation.  In 2009, Liang et al. parameterized a bond-order potential for MoS$_{2}$,\cite{LiangT} which is based on the bond order concept underlying the Brenner potential, \cite{brennerJPCM2002} and where a Lennard-Jones potential is used to describe the weak inter-layer van der Waals interactions in MoS$_{2}$.  This Brenner-like potential was fitted to the phase diagram of Mo and S based materials, and used for MD simulations of friction between MoS$_{2}$ layers. This potential was recently further modified to study the nanoindentation of MoS$_{2}$ thin films using a molecular statics approach. \cite{StewartJA}  A separate force field model was parameterized in 2010 for MD simulations of bulk MoS$_{2}$.\cite{VarshneyV}  

The Tersoff\cite{TersoffJ1,TersoffJ3,TersoffJ2,TersoffJ4,TersoffJ5} and Brenner\cite{brennerJPCM2002} bond-order potentials have had particular success for carbon-based materials.  Beside these two potentials, the Stillinger-Weber (SW) potential is another successful empirical potential for covalently-bonded systems.\cite{StillingerFH} The SW potential has a simpler form and fewer parameters than the Brenner potential, so it is much faster, although it may lose some accuracy in some situations as a trade off. One of the practical advantage of the SW potential is that it has been implemented in almost all available MD simulation packages, where the users' only task is to simply list the SW parameters in the running script. Considering its good translational performance, it is of practical significance to develop a parameterization of the SW potential for SLMoS$_{2}$.

In this paper, we present a parameterization of the SW potential based on fitting the phonon spectrum for SLMoS$_{2}$. There are five interaction terms in this SW potential, which correspond to the five basic modes of motion in SLMoS$_{2}$. This potential provides an accurate prediction for the structural parameters, as well as the Young's modulus of SLMoS$_{2}$. Based on this SW potential, we studied chirality, size, and strain effects on the mechanical and thermal properties of SLMoS$_{2}$ nanoribbons with free edges. The free edges are found to play an important role in impacting both the mechanical and thermal properties of SLMoS$_{2}$.

\begin{figure}[htpb]
  \begin{center}
    \scalebox{0.9}[0.9]{\includegraphics[width=8cm]{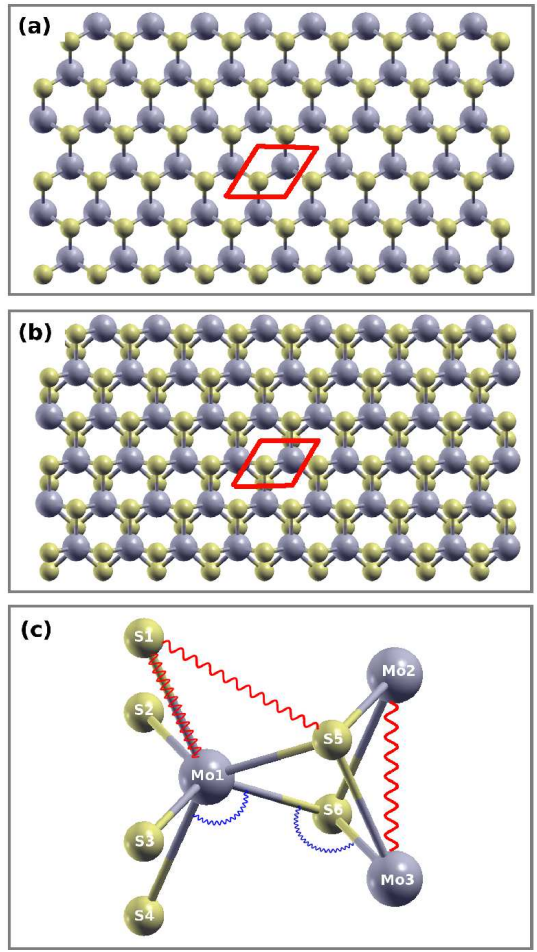}}
  \end{center}
  \caption{(Color online) Configuration of single-layer MoS$_{2}$. (a) Top view. The unit cell is highlighted by a parallelogram (red online). (b) The viewing direction is slightly changed to display the intra-layer structure. (c) Each Mo atom is surrounded by six S atoms, while each S atom is connected to three FNN Mo neighbors. Springs illustrate the five interaction terms: two angle bending terms (blue online) and three bond bending terms (red online).}
  \label{fig_cfg}
\end{figure}

\section{Structure}
Figs.~\ref{fig_cfg}~(a) and (b) show the structure of zigzag SLMoS$_{2}$. Panel (a) is the top view. For clarity, the viewing direction in (b) is changed slightly from (a). The red parallelogram depicts the unit cell in this two-dimensional system, which contains one Mo atom and two S atoms. A SLMoS$_{2}$ includes three atomic layers. The central Mo atomic layer is sandwiched by two S atomic layers with strong intra-layer interaction. We note that we have used the phrase `intra-layer' for these three atomic layers in the SLMoS$_{2}$. We will use another phrase `inter-layer' for the interaction between two MoS$_{2}$ layers. Panel (c) shows a local configuration. Each Mo atom is surrounded by six first-nearest-neighboring (FNN) S atoms, while each S atom is connected to three FNN Mo atoms. The five springs illustrate the five strongest interaction terms. The two blue springs display two kinds of angle bending interactions, while the other three red springs depict the bond bending interactions.

\section{Parametrization of the Stillinger-Weber potential}
\subsection{Covalent Bonding in MoS$_{2}$}
In 1975, Wakabayashi et al. developed a VFF model to analyze their inelastic neutron-scattering data on the phonon spectrum of bulk MoS$_{2}$.\cite{WakabayashiN} This VFF model has been successfully applied to study different MoS$_{2}$-based structures.\cite{JimenezSS,DobardzicE,DamnjanovicM2008mmp} As pointed out by Yu, the VFF model is based on the covalent nature of chemical bonds in the structure.\cite{YuPY}  In the VFF model, five potential terms are introduced for the description of the intra-layer interaction. These terms are essentially equivalent to the five interaction terms depicted by the springs in Fig.~\ref{fig_cfg}. The VFF model is a harmonic potential, and thus is not applicable for the structure relaxation. However, the success of this model implies the atoms in MoS$_{2}$ interact with each other mainly through covalent bonding. This characteristic enables us to develop covalent bonding-based empirical potentials to describe the interatomic interactions for the MoS$_{2}$ system.

\subsection{Stillinger-Weber potential}
As inspired by the VFF model, we now parametrize a SW potential to describe the five interactions shown in Fig.~\ref{fig_cfg}. The SW potential treats the bond bending by a two-body interaction, while the angle bending is described by a three-body interaction. The total potential energy within a system with $N$ atoms is as follows,\cite{StillingerF}
\begin{eqnarray}
\Phi(1,...,N) = \sum_{i<j}V_{2}(i,j) + \sum_{i<j<k}V_{3}(i,j,k).
\end{eqnarray}
The two-body interaction takes following form 
\begin{eqnarray}
V_{2}=\epsilon A\left(B\sigma^{p}r_{ij}^{-p}-\sigma^{q}r_{ij}^{-q}\right)e^{[\sigma\left(r_{ij}-a\sigma\right)^{-1}]},
\label{eq_sw2}
\end{eqnarray}
where the exponential function ensures a smooth decay of the potential to zero at the cut-off, which is key to conserving energy in MD simulations.

The three-body interaction is
\begin{eqnarray}
V_{3}=\epsilon\lambda e^{\left[\gamma\sigma\left(r_{ij}-a\sigma\right)^{-1}+\gamma\sigma\left(r_{jk}-a\sigma\right)^{-1}\right]}\left(\cos\theta_{jik}-\cos\theta_{0}\right)^{2},
\label{eq_sw3}
\end{eqnarray}
where $\theta_{0}$ is the initial angle.

Explicitly, there are five SW potential terms for the SLMoS$_{2}$ corresponding to Fig.~\ref{fig_cfg}~(c). (1) $V_{2}(Mo-S)$ corresponds to the bond bending between the Mo atom and its FNN S atoms. (2) $V_{2}(S-S)$ describes the bond bending between the S atom and its second-nearest-neighboring (SNN) atoms (i.e S atoms). (3) $V_{2}(Mo-Mo)$ is for the bond bending between two SNN Mo atoms. (4) $V_{3}(Mo-S-S)$ describes the bending for the angle with Mo atom as the apex. (5) $V_{3}(S-Mo-Mo)$ is the bending for the angle with an S atom as the apex.

\begin{figure}[htpb]
  \begin{center}
    \scalebox{1.0}[1.0]{\includegraphics[width=8cm]{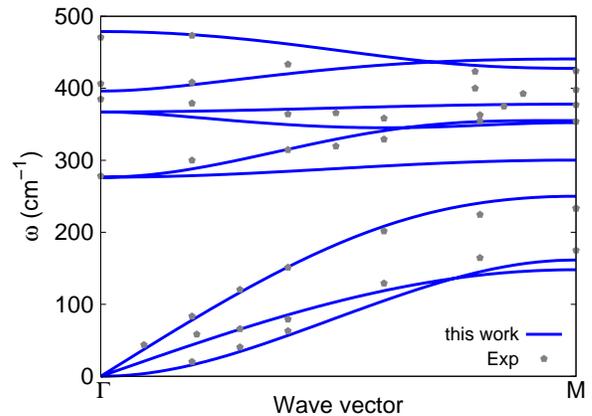}}
  \end{center}
  \caption{(Color online) Phonon spectrum for SLMoS$_{2}$ along the $\Gamma$M direction in the Brillouin zone. The results from the SW potential (blue lines) are fitted to the experiment data (gray pentagons) from Ref~\onlinecite{WakabayashiN}. We note the energy gap around 250~{cm$^{-1}$} and the crossover between the two highest-frequency spectra.}
  \label{fig_phonon_single}
\end{figure}

\begin{figure}[htpb]
  \begin{center}
    \scalebox{1.0}[1.0]{\includegraphics[width=8cm]{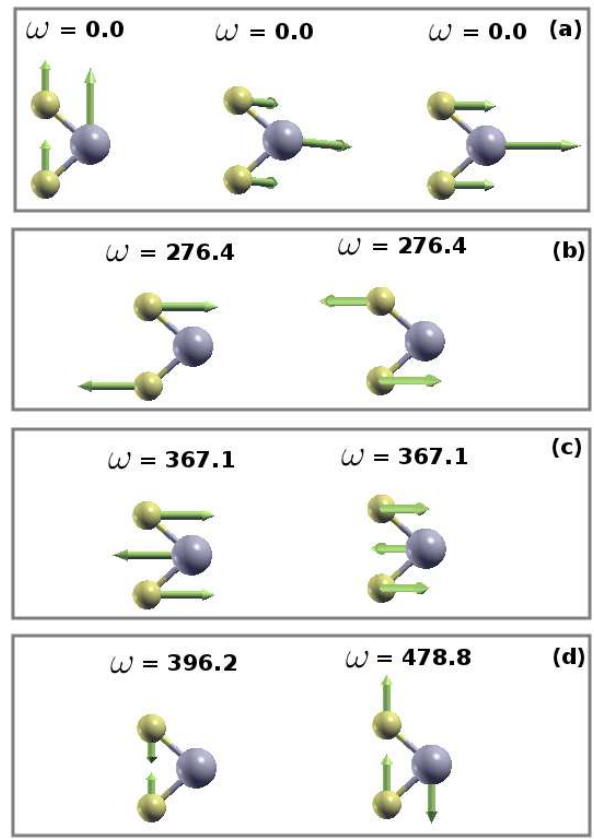}}
  \end{center}
  \caption{(Color online) Eigenvectors for the nine phonon modes at the $\Gamma$ point in SLMoS$_{2}$. (a) Three acoustic phonon modes. (b) Two intra-layer shearing modes, with the two S atomic layers undergoing out-of-phase shearing. (c) Another two intra-layer shearing modes, with the outer two S atomic layers undergoing in-phase shearing. (d) Two intra-layer breathing modes.}
  \label{fig_u}
\end{figure}

\begin{table*}
\caption{The two-body (bond bending) SW potential parameters for GULP.\cite{gulp} The expression is $V_{2}=Ae^{[\rho/\left(r-r_{max}\right)]}\left(B/r^{4}-1\right)$. Energy parameters are in the unit of eV. Length parameters are in the unit of \AA.}
\label{tab_sw2_gulp}
\begin{tabular*}{\textwidth}{@{\extracolsep{\fill}}|c|c|c|c|c|c|}
\hline
& A & $\rho$ & B & $r_{\rm min}$ & $r_{\rm max}$  \\
\hline
Mo-S & 6.0672 & 0.7590 & 15.0787 & 0.0 & 3.16 \\
\hline
S-S & 0.4651 & 0.6501 & 22.3435 & 0.0 & 3.76 \\
\hline
Mo-Mo & 3.5040 & 0.6097 & 25.1333 & 0.0 & 4.27\\
\hline
\end{tabular*}
\end{table*}

\begin{table*}
\caption{Three-body (angle bending) SW potential parameters for GULP.\cite{gulp} The expression is $V_{3}=Ke^{[\rho_{1}/\left(r_{12}-r_{max12}\right)+\rho_{2}/\left(r_{13}-r_{max13}\right)]}\left(\cos\theta-cos\theta_{0}\right)^{2}$.  Energy parameters are in the unit of eV. Length parameters are in the unit of \AA. Mo-S-S indicates the bending energy for the angle with Mo as the apex.}
\label{tab_sw3_gulp}
\begin{tabular*}{\textwidth}{@{\extracolsep{\fill}}|c|c|c|c|c|c|c|c|c|c|c|}
\hline
& K & $\theta_{0}$ & $\rho_{1}$ & $\rho_{2}$ & $r_{\rm min12}$ & $r_{\rm max12}$ & $r_{\rm min13}$ & $r_{\rm max13}$ & $r_{\rm min23}$ & $r_{\rm max23}$ \\
\hline
Mo-S-S & 6.5534 & 81.2279 & 0.6503 & 0.6503 & 0.00 & 3.16 & 0.00 & 3.16 & 0.00 & 3.78   \\
\hline
S-Mo-Mo & 6.5534 & 81.2279 & 0.6503 & 0.6503 & 0.00 & 3.16 & 0.00 & 3.16 & 0.00 & 4.27 \\
\hline
\end{tabular*}
\end{table*}

\begin{table*}
\caption{SW potential parameters for LAMMPS.\cite{Lammps} The two-body potential expression is $V_{2}=\epsilon A\left(B\sigma^{p}r_{ij}^{-p}-\sigma^{q}r_{ij}^{-q}\right)e^{[\sigma\left(r_{ij}-a\sigma\right)^{-1}]}$. The three-body potential expression is $V_{3}=\epsilon\lambda e^{\left[\gamma\sigma\left(r_{ij}-a\sigma\right)^{-1}+\gamma\sigma\left(r_{jk}-a\sigma\right)^{-1}\right]}\left(\cos\theta_{jik}-\cos\theta_{0}\right)^{2}$. tol in the last column is a controlling parameter in LAMMPS. Note that the last two lines only contribute to the two-body (bond bending) interaction, where the three-body interactions are zero ($\lambda=0$). Hence, the other three-body parameters ($\gamma$ and $\cos\theta_{0}$) in the last two lines can be set arbitrarily. Energy parameters are in the unit of eV. Length parameters are in the unit of \AA.}
\label{tab_sw_lammps}
\begin{tabular*}{\textwidth}{@{\extracolsep{\fill}}|c|c|c|c|c|c|c|c|c|c|c|c|}
\hline
& $\epsilon$ & $\sigma$ & a & $\lambda$ & $\gamma$ & $\cos\theta_{0}$ & A & B & p & q & tol \\
\hline
Mo-S-S & 6.0672 & 0.7590 & 4.1634 & 1.0801 & 0.8568 & 0.1525 & 1.0 & 45.4357 & 4 & 0 & 0.0  \\
\hline
S-Mo-Mo & 6.0672 & 0.7590 & 4.1634 & 1.0801 & 0.8568 & 0.1525 & 1.0 & 45.4357 & 4 & 0 & 0.0  \\
\hline
Mo-Mo-Mo & 3.5040 & 0.6097 & 7.0034 & 0.0 & 0.0 & 0.0 & 1.0 & 181.8799 & 4 & 0 & 0.0 \\
\hline
S-S-S & 0.4651 & 0.6501 & 5.7837 & 0.0 & 0.0 & 0.0 & 1.0 & 125.0923 & 4 & 0 & 0.0 \\
\hline
\end{tabular*}
\end{table*}

The SW potential has succeeded in many covalent bonding systems, such as diamond-like structures,\cite{StillingerF} graphene,\cite{AbrahamFF} etc. There are several advantages in the SW potential. (1) First, each potential term carries a clear physical interpretation. The two-body term is due to the bond bending interaction, and the three-body term captures the angle bending movement. This advantage is inherited from the VFF model. (2) Secondly, the anharmonic portion in the SW potential has provided reasonably accurate information for nonlinear effects in covalent systems.\cite{AbrahamFF} This nonlinear property is important for MD simulations of nonlinear phenomena like thermal conductivity. (3) Finally, there are several practical advantages of the SW potential. The most important one is the wide application of this SW potential. It has been implemented in most available lattice dynamics or molecular dynamics simulation packages, such as GULP,\cite{gulp} LAMMPS,\cite{Lammps} etc. The SW potential parameterized in present work can be applied directly in these simulation packages, by simply changing the SW parameters in the running script.

\subsection{SW parameterization}
The code GULP\cite{gulp} was used to calculate the fitting parameters for the SW potential, where the SW parameters were fit to the phonon spectrum of SLMoS$_{2}$.  The phonons are the most important heat energy carrier in the thermal transport in semiconductors. Thus the phonon spectrum plays a crucial role for the thermal conductivity. Furthermore, the acoustic phonon velocities from the phonon spectrum are closely related to the mechanical properties of the material.\cite{LandauLD} As a result, a good fitting to the phonon spectrum will automatically lead to a good description for mechanical properties.

Fig.~\ref{fig_phonon_single} shows the fitting results for the phonon spectrum of SLMoS$_{2}$ along the $\Gamma$M line in the Brillouin zone. The SW potential is fitted for SLMoS$_{2}$ (not bulk MoS$_{2}$). This can greatly simplify the fitting procedure as compared to fitting for bulk MoS$_{2}$, because it is more difficult to fit the potential to two nearly degenerate curves. The experimental data is from Ref~\onlinecite{WakabayashiN}. \jwj{It has been shown that the phonon spectrum from the first-principles calculation agrees with these experimental data.\cite{SanchezAM} It means that our theoretical results on the phonon spectrum also agrees with the first-principles calculations.}  We note that while the experimental data is for bulk MoS$_{2}$, we have adopted the data for SLMoS$_{2}$. We do so by reading out the experimental frequency for the intra-layer phonon modes from one of the two almost degenerate curves. The inter-layer phonon modes in bulk MoS$_{2}$ are determined by the inter-layer van der Waals interaction, so they are not considered here.  This is justifiable, however, because the only difference between bulk MoS$_{2}$ and SLMoS$_{2}$ is the weak van der Waals interaction between adjacent S atomic layers.\cite{LiangT} The van der Waals interaction is substantially weaker than the strong intra-layer SW interactions, which will be confirmed later.  We will show later that the resulting SW potential can be combined with the van der Waals interaction to describe the interaction within bulk MoS$_{2}$.

The phonon spectrum in Fig.~\ref{fig_phonon_single} is calculated with the SW parameters shown in Tabs.~\ref{tab_sw2_gulp} and ~\ref{tab_sw3_gulp}. The SW potential has been slightly reformulated in GULP as shown in the caption of the two tables. All cut-offs are determined from the structure of SLMoS$_{2}$ in Fig.~\ref{fig_cfg}~(c). These values are not relaxed during the fitting procedure. We note that GULP has introduced an additional cut-off ($r_{\rm max23}$) for the angle bending potential terms $V_{3}(Mo-S-S)$ and $V_{3}(S-Mo-Mo)$. This additional cut-off is critical if the apex atom has more than 4 FNN atoms. For example, the additional cut-off is very important for the $V_{3}(Mo-S-S)$ potential term. Specifically, in Fig.~\ref{fig_cfg}~(c), there are six FNN atoms around the Mo$_{1}$ atom: S$_{1}$, S$_{2}$, S$_{3}$, S$_{4}$, S$_{5}$, and S$_{6}$. These atoms form two different types of angles: $\angle S_{4}Mo_{1}S_{6}$ and $\angle S_{4}Mo_{1}S_{5}$. The length $r_{S4-S6}$ is different from $r_{S4-S5}$, so these two angles can be distinguished by the additional cut-off. However, they become indistinguishable if the additional cut-off is not introduced.  We wish to consider interactions only for the bending of the first angle $\angle S_{4}Mo_{1}S_{6}$, which is why we state that the additional cut-off in GULP plays a critical role.

Several basic physical considerations have been found to be helpful during the fitting procedure.  First of all, the initial input SW parameters are estimated from different interaction terms in the VFF model, which provides a reasonable basic parameter set for the fitting procedure. Furthermore, each of the five interaction terms dominates the corresponding phonon branches in the phonon spectrum. To disclose the mapping between the interaction terms and the phonon branches, it is helpful to consider the eigenvector of each phonon branch. Fig.~\ref{fig_u} shows the motion (eigenvectors) for the nine phonon modes at the $\Gamma$ point. The figure is plotted with XCRYSDEN.\cite{xcrysden} Panel (a) shows the three translational acoustic modes. Their frequencies are zero due to the rigid translational symmetry in the SW potential.\cite{BornM} These three branches get a major contribution from the $V_{2}(Mo-S)$, $V_{3}(Mo-S-S)$, and $V_{3}(S-Mo-Mo)$ terms. Panel (b) shows two intra-layer shearing modes, which have the same frequency. The two outer S atomic layers involve in an out-of-phase shearing motion, while the inner Mo atomic layer is stationary. Obviously, this motion is governed by the two angle bending terms $V_{3}(Mo-S-S)$ and $V_{3}(S-Mo-Mo)$. Panel (c) shows another two intra-layer shearing modes. The movement of the two S atomic layers is in-phase. This motion style can be regarded as if the two S atomic layers are still, while the inner Mo atomic layer is oscillating. This motion style is controlled mainly by the two angle bending terms $V_{3}(Mo-S-S)$ and $V_{3}(S-Mo-Mo)$. These four intra-layer shearing modes in panels (b) and (c) become dispersed at the edge of the Brillouin zone. This dispersion is controlled by the $V_{2}(Mo-Mo)$ and $V_{2}(S-S)$ potential terms. Panel (d) shows two intra-layer breathing modes. In the first mode, the two S atomic layers oscillate out-of-phase, so the $V_{2}(S-S)$ potential term makes a significant contribution for this mode. We have taken advantage of the mapping between the interaction terms and the phonon branches during the fitting procedure.

As another verification of our SW potential, we note the crossover between the two highest-frequency optical branches. This crossover also occurs in first-principles calculation results,\cite{SanchezAM} yet it is absent in the VFF model.\cite{WakabayashiN} These two branches correspond to the two phonon modes in Fig.~\ref{fig_u}. From the motion of these two modes, it can be found that a smaller $V_{2}(S-S)$ interaction between the two outer S atomic layers is important for this crossover to occur. If this interaction between the two S atomic layers is too strong, then the first mode (with S layers oscillating out-of-phase) will have a larger frequency than the second mode (with S layers oscillating in-phase). In this case, the two branches will diverge from each other instead of crossing over. In the VFF model,\cite{WakabayashiN} the S-S coupling is on the same order as the coupling between Mo-Mo, so the crossover is missing. In our calculation, the S-S interaction (0.4651~{eV}) is much smaller than the Mo-Mo interaction (3.5040~{eV}), which enables us to capture the crossover phenomenon. Our SW potential fit has also preserved the energy gap around 250~{cm$^{-1}$} between the acoustic and optical branches. This is made possible from a proper choice of the strength ratio between the two-body and three-body interactions.

Tab.~\ref{tab_sw_lammps} shows the SW parameters used in LAMMPS. The expression of the SW potential used in LAMMPS is shown in the caption, which is exactly the same as the original SW paper.\cite{StillingerF}. All values in this table are derived from Tabs.~\ref{tab_sw2_gulp} and ~\ref{tab_sw3_gulp} by comparing their expressions of the SW potential. The potential script for this SW potential in LAMMPS can be found in the supplemental material.\cite{supplemental} As long as LAMMPS uses exactly the original expression for the SW potential, there is no any additional cut-off for the three-body interaction. As a result, LAMMPS cannot distinguish $\angle S_{4}Mo_{1}S_{6}$ and $\angle S_{4}Mo_{1}S_{5}$, so the $V_{3}(Mo-S-S)$ potential term is applied to both angles. This makes simulation impossible, because these two angles are quite different and should not have the same interaction. Furthermore, we hope to consider interactions only for $\angle S_{4}Mo_{1}S_{6}$, not for $\angle S_{4}Mo_{1}S_{5}$. We thus have made an ad hoc modification in  the source file $pair\_sw.cpp$ to embed the additional cut-off. The modified source file $pair\_sw.cpp$ can be found in the supplemental material.\cite{supplemental} Users can recompile their LAMMPS package using the modified $pair\_sw.cpp$ file, and then start their simulations on SLMoS$_{2}$ with the SW potential script in the supplemental material.\cite{supplemental}

\subsection{Discussion on the SW potential}

\begin{table}
\caption{The Lennard-Jones potential parameters for the inter-layer coupling between two neighboring S atomic layers from Ref.~\onlinecite{LiangT}. The expression is $V=4\epsilon \left[\left(\frac{\sigma}{r}\right)^{12} - \left(\frac{\sigma}{r}\right)^{6}\right]$. The cut-off is 10.0~{\AA}.}
\label{tab_lj}
\begin{tabular}{@{\extracolsep{\fill}}|c|c|}
\hline
$\epsilon$ (eV) & $\sigma$ (\AA)  \\
\hline
0.00693 & 3.13 \\
\hline
\end{tabular}
\end{table}

\begin{figure}[htpb]
  \begin{center}
    \scalebox{1.0}[1.0]{\includegraphics[width=8cm]{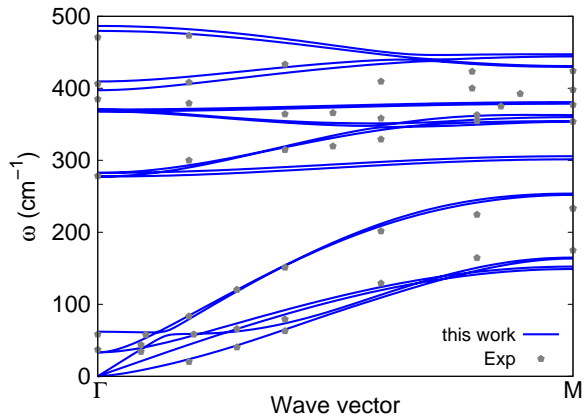}}
  \end{center}
  \caption{(Color online) Phonon spectrum for bulk MoS$_{2}$ along the $\Gamma$M direction in the Brillouin zone. The results from the SW potential (blue lines) are compared with the experiment data (gray pentagons) from Ref~\onlinecite{WakabayashiN}. We note the agreement between the SW results and the experiment for the two low-frequency optical modes at the $\Gamma$ point (inter-layer shearing and breathing modes).}
  \label{fig_phonon_bulk}
\end{figure}

\begin{figure}[htpb]
  \begin{center}
    \scalebox{1.0}[1.0]{\includegraphics[width=8cm]{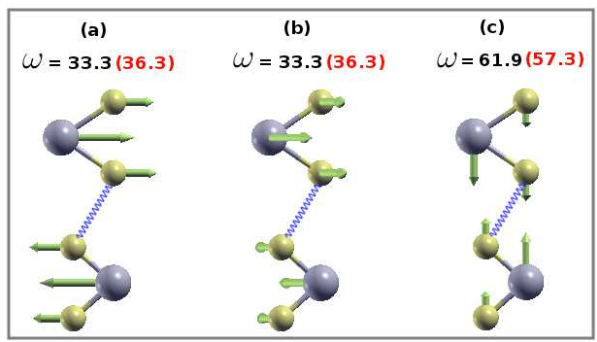}}
  \end{center}
  \caption{(Color online) The eigenvectors for the two inter-layer shearing modes and the inter-layer breathing mode. (a) and (b) are the two shearing modes with degenerate frequency. (c) The inter-layer breathing mode. Numbers are the frequency calculated from the SW potential (black online), which is compared with the experiment data\cite{WakabayashiN} in parentheses (red online).}
  \label{fig_u_bulk}
\end{figure}

\begin{figure}[htpb]
  \begin{center}
    \scalebox{1.0}[1.0]{\includegraphics[width=8cm]{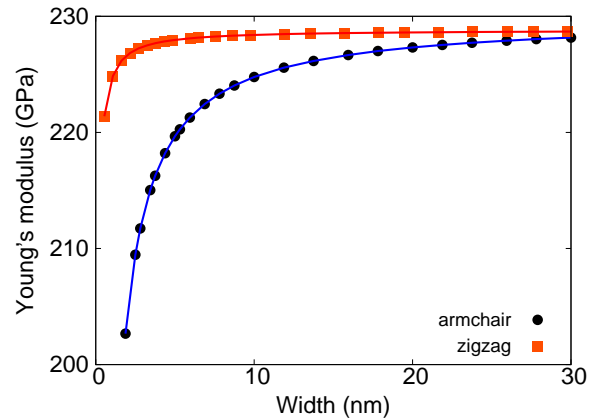}}
  \end{center}
  \caption{(Color online) Size-dependent Young's modulus for armchair and zigzag SLMoS$_{2}$ nanoribbons (i.e. with free boundary conditions). The length of the SLMoS$_{2}$ is 5~nm. Both curves converge to the same value of 229.0~{GPa}, which coincides with the Young's modulus of SLMoS$_{2}$ with PBCs.}
  \label{fig_young}
\end{figure}

As we have pointed out in the discussion for Fig.~\ref{fig_phonon_single}, the experimental data used for the fitting procedure are actually for bulk MoS$_{2}$ instead of SLMoS$_{2}$. For bulk MoS$_{2}$, the adjacent MoS$_{2}$ layers are weakly coupled through van der Waals interactions, which can be well described by the Lennard-Jones potential. Tab.~\ref{tab_lj} lists the two parameters for the Lennard-Jones potential as developed in Ref.~\onlinecite{LiangT}. The SW potential combined with this Lennard-Jones potential enables us to simulate bulk or few-layer MoS$_{2}$.

Fig.~\ref{fig_phonon_bulk} shows the phonon spectrum in bulk MoS$_{2}$ along the $\Gamma$M high symmetric line in the Brillouin zone. Our theoretical results agree quite well with the experimental data for bulk MoS$_{2}$. Each curve in the SLMoS$_{2}$ is split into two closing curves in bulk MoS$_{2}$, due to the weak inter-layer van der Waals interactions. A distinct influence of the inter-layer van der Waals coupling is the generation of three low-frequency branches around 30~{cm$^{-1}$} and 60~{cm$^{-1}$}. Fig.~\ref{fig_u_bulk} shows the eigenvectors of these three phonon modes at the $\Gamma$ point. (a) and (b) are two inter-layer shearing modes. (c) is the inter-layer breathing mode. It is important to note that the calculated frequencies for both inter-layer shearing and breathing modes are in good agreement with the experimental data (red numbers).

\section{Structural Characterization and Bulk and Size-Dependent Elastic Properties}

To characterize the structural parameters of SLMoS$_{2}$ in the relaxed configuration, the Mo-S bond length from our calculation is 2.3920~{\AA}, which is quite close to 2.382~{\AA} from first-principles calculations.\cite{SanchezAM} The intra-layer lattice constant is 3.0937~{\AA}, which agrees with the experimental value of 3.15~{\AA}.\cite{WakabayashiN} For bulk MoS$_{2}$, the vertical lattice constant is 12.184394~{\AA}, which is also close to the experimental value 12.3~{\AA}.\cite{WakabayashiN} The angles are $\angle SMoS=\angle MoSMo=80.581^{\circ}$.

The SW potential has been fitted to the phonon spectrum as discussed above. The acoustic phonon branches are related to the in-plane Young's modulus. Hence, it is natural to expect that the SW potential developed above should also give reasonable prediction for the Young's modulus of SLMoS$_{2}$. We find that the Young's modulus for SLMoS$_{2}$ is 229.0~{GPa} if periodic boundary conditions (PBCs) are applied in the two intra-plane directions, which does not depend on the chirality or the size of the system. To extract the value of the Young's modulus, we have adopted 6.092197~{\AA} as the thickness of SLMoS$_{2}$, which is half of the vertical lattice constant.  Recent experiments have measured the effective Young's modulus to be $E=120\pm 30$~{Nm$^{-1}$},\cite{CooperRC2013prb1,CooperRC2013prb2} or $E=180\pm 60$~{Nm$^{-1}$}.\cite{BertolazziS} These values correspond to an in-plane Young's modulus of $198.6\pm 49.7$~{GPa} or $297.9\pm 99.3$~{GPa}, considering the thickness of 6.092197~{\AA}. Our theoretical value of 229.0 GPa coincides with the experimental value within its error. 

If free boundary conditions (FBCs) within the two-dimensional plane of SLMoS$_{2}$ are applied, then so-called edge effects become active and important, as has previously been observed for graphene.\cite{KimSY2009nl,JiangJW2009edge,ReddyCD,JiangJW2012jap} Edge effects arise from the fact that atoms at the boundaries of two-dimensional crystals are undercoordinated, i.e. have fewer bonding neighbors than atoms within the crystal bulk, and thus edge effects are the two-dimensional analog of surface effects for three-dimensional crystals.  Their effects on the elastic properties\cite{ShenoyVB,ReddyCD} and the mechanical quality factors\cite{KimSY2009nl,JiangJW2012jap} of graphene have recently been investigated.  Fig.~\ref{fig_young} shows that the Young's modulus is sensitive to both the chirality and width of SLMoS$_{2}$. In this calculation, the length is kept constant at 5~{nm} and the results are found to be independent of the length. Interestingly, the Young's modulus for both zigzag and armchair SLMoS$_{2}$ decreases as the width decreases. This is different from the size-dependent trend observed by Reddy et al\cite{ReddyCD}, but similar as the results found by Zhao et al.\cite{ZhaoH2009nl} for graphene nanoribbons.  The Young's modulus is smaller for armchair SLMoS$_{2}$ than zigzag SLMoS$_{2}$, and both curves converge to the same value as in the PBC calculation, i.e. 229.0~{GPa}, which corresponds to SLMoS$_{2}$ without edge effects, as the width increases.  

\section{Thermal conductivity of SLMoS$_{2}$}
In the previous sections, we have developed a SW potential to describe the interactions within SLMoS$_{2}$. The remainder of this paper focuses on simulations of thermal transport in both infinite, periodic 
SLMoS$_{2}$, and SLMoS$_{2}$ nanoribbons with free edges, whose physical properties are different due to the edge effects.

\subsection{Simulation details}

\begin{figure}[htpb]
  \begin{center}
    \scalebox{1.0}[1.0]{\includegraphics[width=8cm]{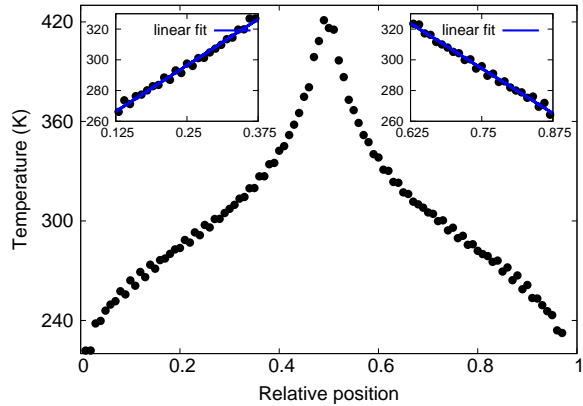}}
  \end{center}
  \caption{(Color online) Temperature profile at 300~K for armchair SLMoS$_{2}$ with PBC. The current ratio $\alpha=0.1$ and the relaxation time of the heat bath is $\tau=0.04$~ps. The top left inset shows the linear fitting for the profile in $x\in[0.125, 0.375]$, giving a temperature gradient $\frac{dT}{dx}_{1}$. The right inset shows the linear fitting for the profile in $x\in[0.625, 0.875]$, giving a temperature gradient $\frac{dT}{dx}_{2}$. These two temperature gradients ($\frac{dT}{dx}_{1}$ and $\frac{dT}{dx}_{2}$) are averaged in the calculation of the thermal conductivity using the Fourier law.}
  \label{fig_dTdx}
\end{figure}

\begin{figure}[htpb]
  \begin{center}
    \scalebox{0.9}[0.9]{\includegraphics[width=8cm]{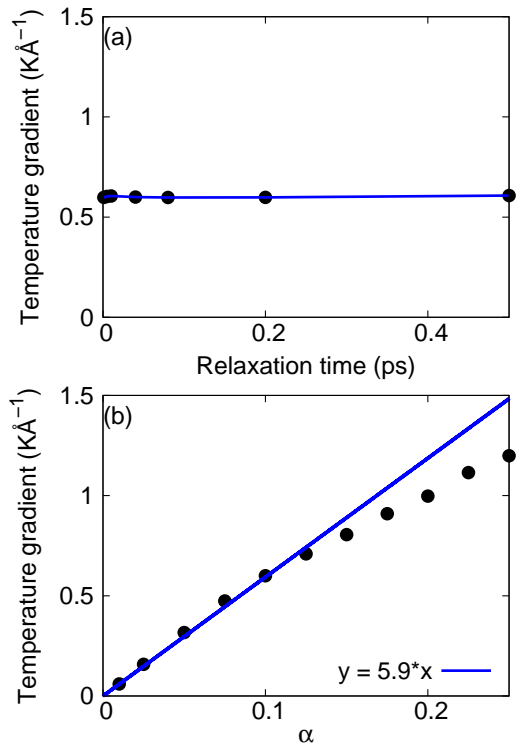}}
  \end{center}
  \caption{(Color online) Temperature gradients from the calculation with different simulation parameters in armchair SLMoS$_{2}$ with PBC. (a) The relaxation time $\tau$ has almost no effect on the simulation result. The current ratio $\alpha=0.1$. (b) The temperature gradient increases linearly with increasing current ratio in small $\alpha$ region. The increasing deviates obviously from linear for $\alpha>0.15$. The relaxation time $\tau=0.04$~{ps}.}
  \label{fig_alpha_tau}
\end{figure}

\begin{figure*}[htpb]
  \begin{center}
    \scalebox{0.9}[0.9]{\includegraphics[width=\textwidth]{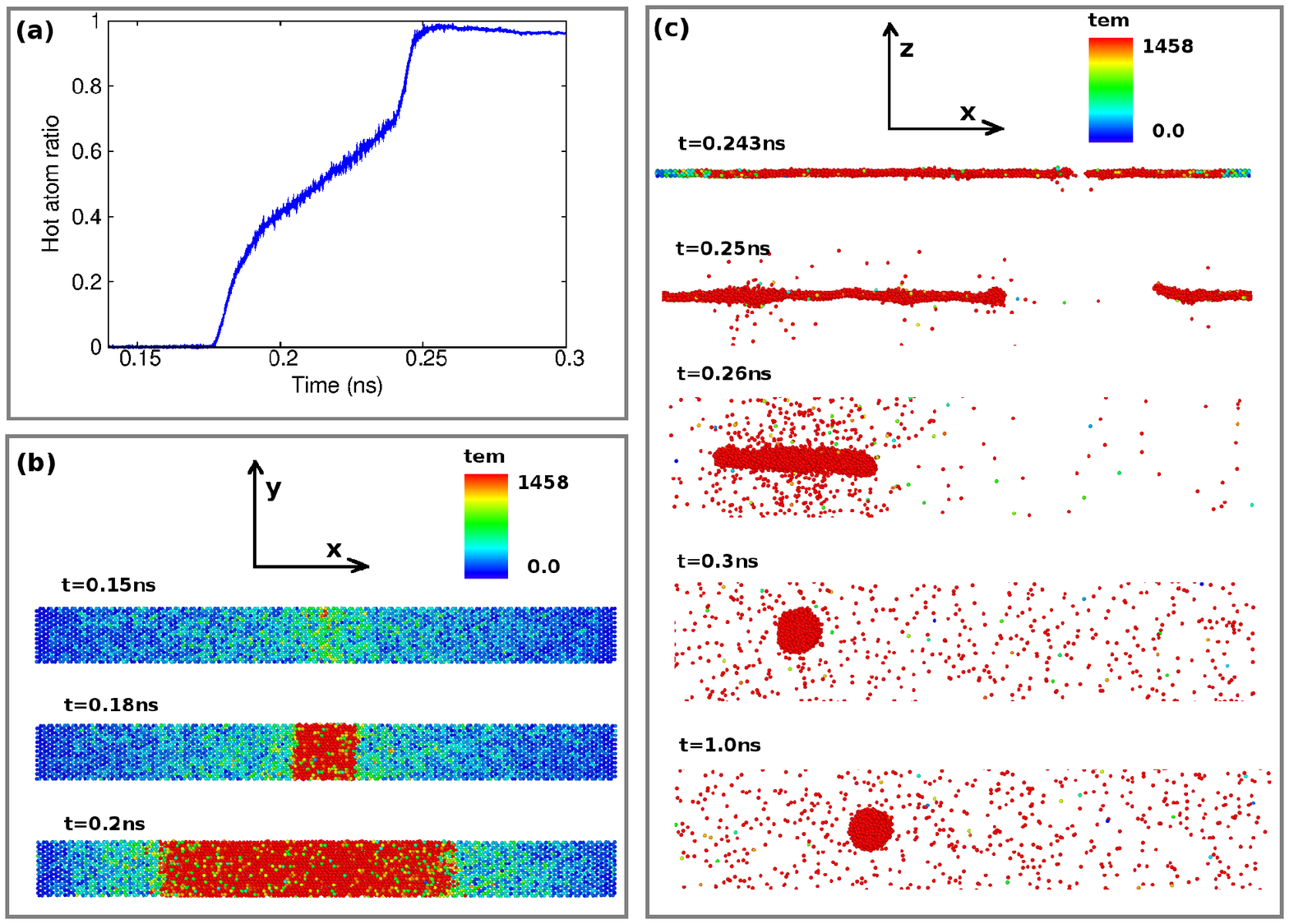}}
  \end{center}
  \caption{(Color online) Melting of armchair SLMoS$_{2}$ with PBCs at 300~K with a large current ratio of $\alpha=0.3$. (a) The hot atom number to the total atom number ratio. Hot atoms are those with temperature above the melting point $T_{m}=1458$~K. The melting process begins around 0.16~ns and ends quickly. A large current ratio leads to high temperature in the hot temperature-controlled region, resulting in higher melting possibility. (b) Snap-shots for initial stages of the melting process. The melting starts from some atoms in the hot temperature-controlled region, and propagates to the two cold temperature-controlled regions. (c) Snap-shots for the final stages of the melting process. The SLMoS$_{2}$ fractures, followed by evaporation of most of the atoms.  Finally, the remaining atoms aggregate into a spherical nano-particle. The size of the resulting nano-particle is stabilized through exchanging surface atoms with the evaporated atoms. Color in (b) and (c) is with respect to the temperature.}
  \label{fig_melt}
\end{figure*}

The thermal transport is simulated by the MD simulation method implemented in the LAMMPS package.\cite{Lammps} The thermal current across the system is driven by frequently moving kinetic energy from the cold region to the hot region.\cite{IkeshojiT} This energy transfer is accomplished via scaling the velocity of atoms in the hot or cold temperature-controlled regions. The total thermal current pumped into the hot region (or pumped out from the cold region) is $J = \alpha E_{k}^{\rm hot}$. We have introduced a current parameter $\alpha$ to measure the energy amount to be aggregated per unit time. $E_{k}^{\rm hot} = N_{\rm hot} * 1.5k_{B}T$ is the statistical total kinetic energy in the hot region, where $N_{\rm hot}$ is the number of atoms in the hot temperature-controlled region. $k_{B}$ is the Boltzmann constant, and $T$ is the temperature. We note that the total current $J$ is the \textit{eflux} parameter in the \textit{fix heat} command in LAMMPS. We point out the physical meaning of the current parameter $\alpha$. This parameter is the ratio between the aggregated energy and the kinetic energy of each atom, i.e $\alpha$ measures the thermal current for each atom with respect to its kinetic energy. A larger current parameter leads to a larger temperature gradient. For example, if $\alpha=1.0$, then atoms in the cold region will be frozen, because their kinetic energies are fully removed to the hot region. The velocity scaling operation is performed every $\tau$ ps, which can be regarded as the relaxation time of this particular heat bath.

After thermalization for a sufficiently long time, a temperature profile is established across the SLMoS$_{2}$. Fig.~\ref{fig_dTdx} shows the temperature profile for armchair SLMoS$_{2}$ with size $39.51\times3.75\times0.60$~{nm}, which has PBCs in both in-plane directions. The central region is put in the hot temperature-controlled region, and the two ends are put in the cold temperature-controlled region. The two insets show that the left middle region with $x\in[0.125, 0.375]$ and the right middle region with $x\in[0.625, 0.875]$ are linearly fitted to extract two temperature gradients. These two temperature gradients are averaged to give the final temperature gradient $\frac{dT}{dx}$. The thermal conductivity of the SLMoS$_{2}$ is obtained through the Fourier law,
\begin{eqnarray}
\kappa = \frac{1}{2} \times \frac{J}{|dT/dx|},
\label{eq_fourier}
\end{eqnarray}
where a factor of $\frac{1}{2}$ comes from the fact that heat current flows in two opposite directions.

\subsection{Simulation results and discussions}
We first examine the effects of the two parameters for the thermal transport simulation, i.e the relaxation time $\tau$ and the current parameter $\alpha$. We simulate the thermal transport at 300~K in armchair SLMoS$_{2}$ with different $\tau$ and $\alpha$.  PBCs have been applied in this case to eliminate edge effects. Fig.~\ref{fig_alpha_tau}~(a) shows that the temperature gradient is not sensitive to the relaxation time. The current parameter $\alpha =0.1$ in these simulations, so the thermal current is also unchanged for these simulations. Hence, the obtained thermal conductivity is not sensitive to the relaxation time. We choose $\tau=0.04$~{ps} in all following simulations.

Fig.~\ref{fig_alpha_tau}~(b) shows the effect from the current parameter $\alpha$.  From its definition, the thermal current $J$ increases linearly with increasing current parameter $\alpha$. The figure shows that the temperature gradient also increases linearly with increasing current parameter for $\alpha\le 0.125$, which means that the obtained thermal conductivity does not depend on the current parameter value for $\alpha\le 0.125$. However, the increase of the temperature gradient clearly deviates from the linear behavior for $\alpha\ge 0.15$. As a result, one gets different values for the thermal conductivity for different current parameters $\alpha$, which is artificial. It actually indicates the violation of the Fourier law in case of large temperature gradient, because the Fourier law is a linear phenomenological law.

If the current parameter $\alpha$ is large, then the atoms in the hot region are more likely to reach a high temperature, which can lead to melting of SLMoS$_{2}$ from the hot region, considering that its melting point is about 1458~K.\cite{ZhangX1999cc} Fig.~\ref{fig_melt} illustrates the melting process of armchair SLMoS$_{2}$ at 300~K with a large current parameter of $\alpha=0.3$. The snap-shots are produced by OVITO.\cite{ovito} Fig.~\ref{fig_melt}~(a) shows the hot atom number to the total atom number ratio, where a hot atom is defined to be an atom with temperature above the melting point. The melting process begins around 0.16~ns and is finished quickly within 90~{ps}. Fig.~\ref{fig_melt}~(b) are snap-shots for the initial melting process. The melting starts from some atoms in the hot temperature-controlled region, and propagates to the two cold temperature-controlled regions. Fig.~\ref{fig_melt}~(c) are snap-shots for the final stages of the melting process, which shows that the armchair SLMoS$_{2}$ has fractured, leading to evaporation of most of the atoms, while the remaining atoms aggregate into a spherical nano-particle. The size of the resulting nano-particle is stabilized through exchanging surface atoms with these atoms that have been evaporated into the air.

\begin{figure}[htpb]
  \begin{center}
    \scalebox{1.0}[1.0]{\includegraphics[width=8cm]{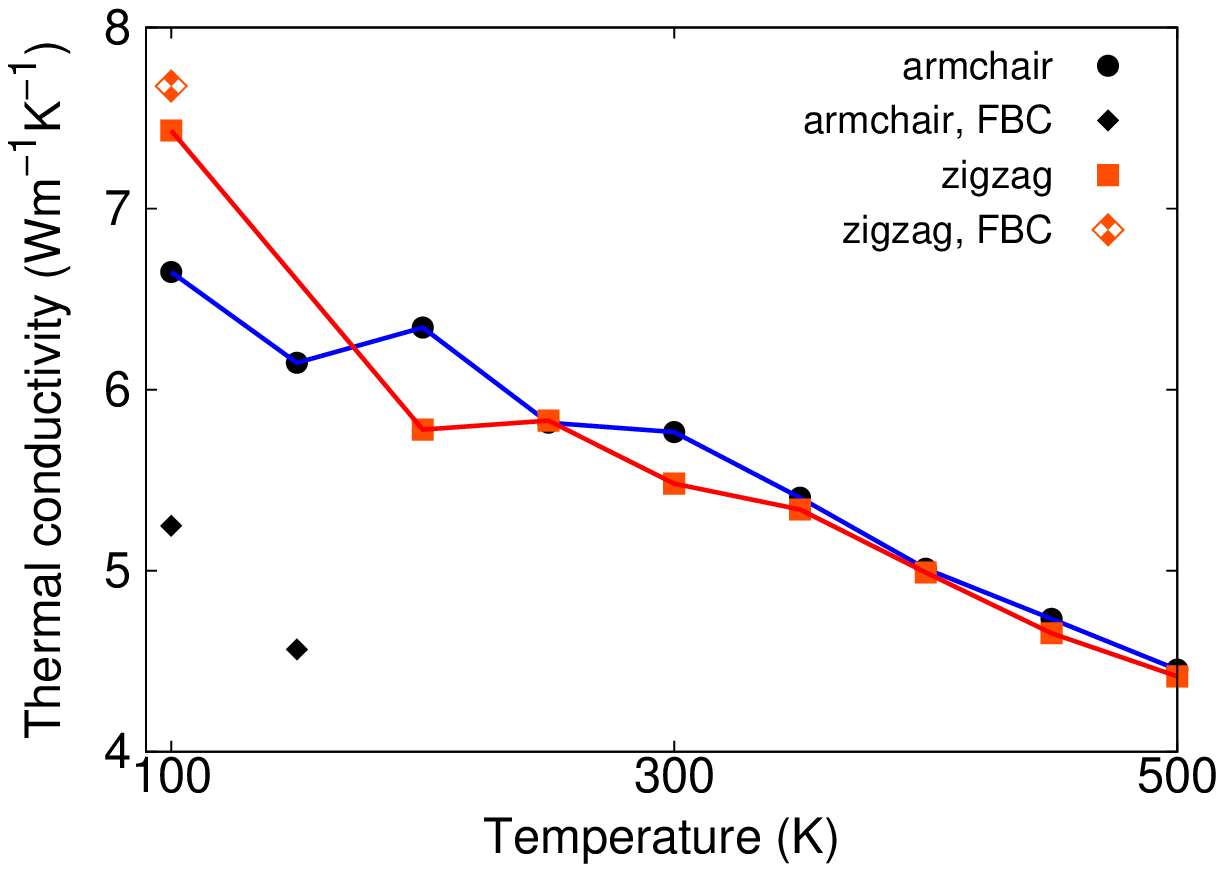}}
  \end{center}
  \caption{(Color online) Temperature dependence for the thermal conductivity. The chirality does not appear to significantly effect the thermal conductivity. The armchair MoS$_{2}$ nanoribbon with free edges (FBCs) has a much lower thermal conductivity than that without free edges.}
  \label{fig_conductivity_temperature}
\end{figure}

\begin{figure}[htpb]
  \begin{center}
    \scalebox{0.9}[0.9]{\includegraphics[width=8cm]{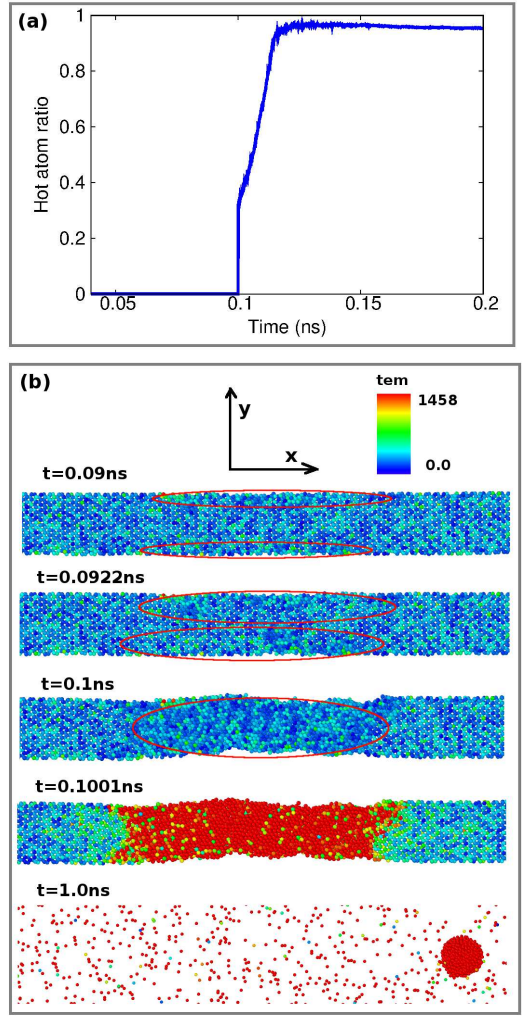}}
  \end{center}
  \caption{(Color online) The melting process in the armchair SLMoS$_{2}$ with free edges at 300~K. (a) The hot atom number to the total atom number ratio. The whole melting process is completed rapidly. (b) The melting process starts from the free edge, which possess some edge modes as shown in Fig. The edge modes cause serve damage to the free edges, inducing the induction of the melting phenomenon. An obvious nano-particle is also generated as the outcome of the melting phenomenon.}
  \label{fig_melt_fbc}
\end{figure}

\begin{figure}[htpb]
  \begin{center}
    \scalebox{1.0}[1.0]{\includegraphics[width=8cm]{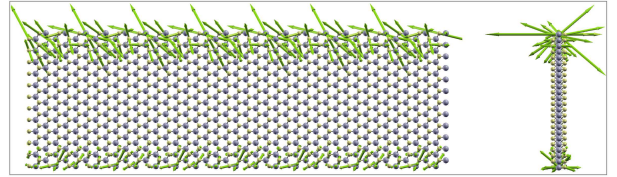}}
  \end{center}
  \caption{(Color online) One edge mode in the armchair SLMoS$_{2}$ with free edges. Left and right are the top and side views, respectively.}
  \label{fig_edgemode}
\end{figure}

\begin{figure}[htpb]
  \begin{center}
    \scalebox{1.0}[1.0]{\includegraphics[width=8cm]{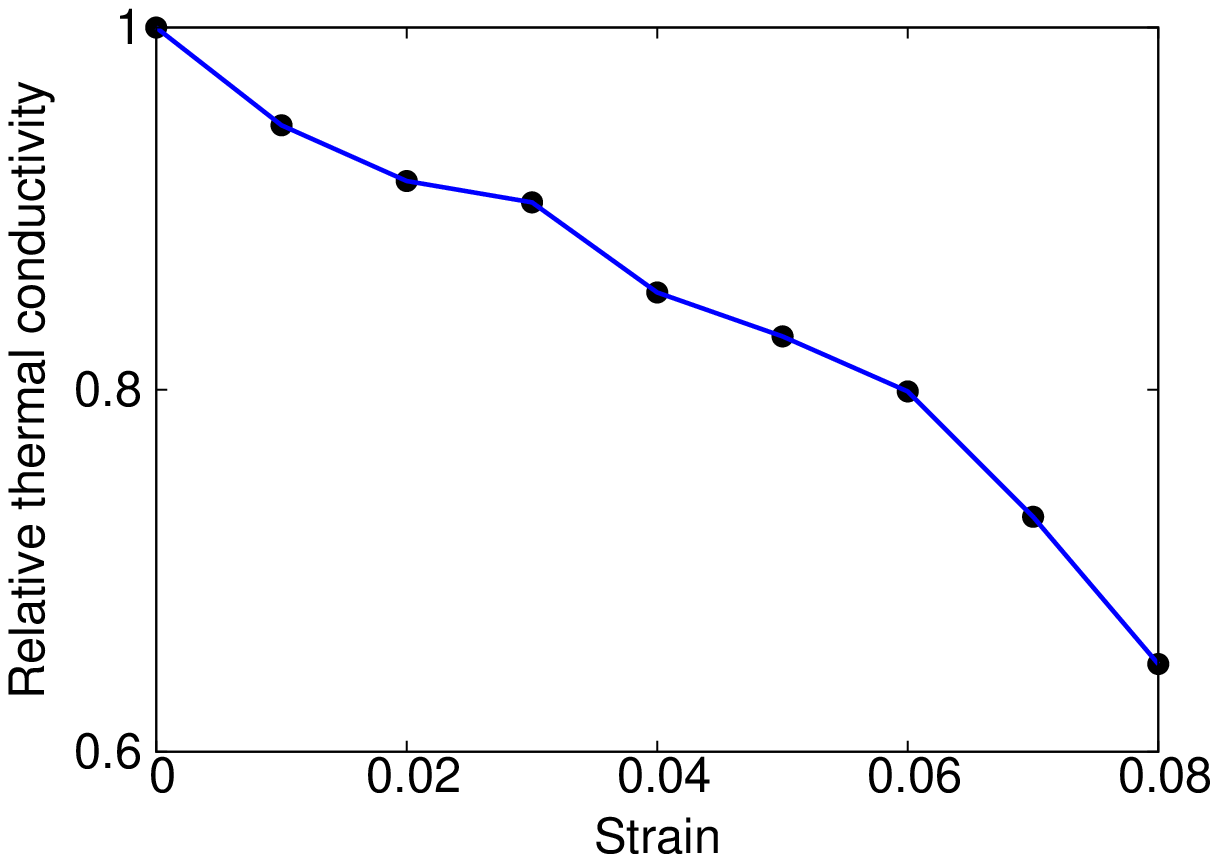}}
  \end{center}
  \caption{(Color online) Strain effect on the thermal conductivity at 300~K of the armchair SLMoS$_{2}$ with PBC.}
  \label{fig_conductivity_strain}
\end{figure}

Fig.~\ref{fig_conductivity_temperature} shows the temperature dependence for the thermal conductivity of SLMoS$_{2}$, where $\alpha=0.1$ for all simulations. The zigzag SLMoS$_{2}$ has the size $40.0\times 3.88\times 0.60$~{nm}, which is almost the same as the armchair SLMoS$_{2}$. There is no obvious difference in the thermal conductivity between the armchair and zigzag SLMoS$_{2}$ with PBCs in the in-plane lateral direction. This is the result of the three-fold rotational symmetry in SLMoS$_{2}$. This symmetry requires all physical properties, which are second-order tensors, to be isotropic in all in-plane directions.\cite{BornM} Due to this requirement, the thermal conductivity obtained from the Fourier law is isotropic.

Fig.~\ref{fig_conductivity_temperature} also shows that the free edges have an important effect on the thermal conductivity. The armchair SLMoS$_{2}$ with FBCs in the in-plane lateral direction has much lower thermal conductivity than that with PBC at 100 and 150~K. For the zigzag SLMoS$_{2}$, the thermal conductivity at 100~K is almost the same with either PBC or FBC. The thermal conductivity at higher temperature for SLMoS$_{2}$ with FBC are not shown in the figure, because the free edges in the SLMoS$_{2}$ are not stable at higher temperature, as will be discussed next.

Fig.~\ref{fig_melt_fbc} shows that the free edges can induce a melting phenomenon in SLMoS$_{2}$ at 300~K with $\alpha=0.1$. Panel (a) shows the hot atom ratio, where the melting process starts around 0.1~ns and finishes rapidly within 15~ps. Snap-shots in Fig.~\ref{fig_melt_fbc}~(b) show that the melting initiates at the free edges. At t = 0.09~ns, there is obvious disorder at the two free edges, which are indicated by red ellipses. These disordered regions propagate into the center and finally merge into a large disordered region. Actually, these disordered regions are amorphous structures, which occurs due to the lack of constraint within the two-dimensional plane because of the free edges and dangling edge bonds.  Subsequent melting of the entire SLMoS$_{2}$ occurs rapidly. Finally, a nanoparticle is also generated as the outcome of the melting phenomenon. 

The initial disordered region at the free edges is due to the localization of edge modes in these regions. Fig.~\ref{fig_edgemode} shows a typical edge mode in the SLMoS$_{2}$ with free edges. In this mode, only edge atoms vibrate with large amplitude, while the inner atoms exhibit comparably low amplitude motion. As a result, the edge regions have higher possibility to form defects. This free edge induced melting phenomenon is different from the temperature-induced melting for system without free edges as shown in Fig.~\ref{fig_melt}.  Specifically, the free edge induced melt occurs at lower temperature and with a smaller current parameter than the temperature-induced melt. For the given current parameter $\alpha=0.1$ which means that the Fourier law is obeyed, the melting temperature for infinite, periodic (i.e. with PBCs in-plane) armchair or zigzag SLMoS$_{2}$ is about 600~K. However, in the presence of free edges, this melting temperature drops to about 200~K or 150~K for armchair or zigzag SLMoS$_{2}$, respectively.

Finally, Fig.~\ref{fig_conductivity_strain} presents the tensile strain effect on the thermal conductivity at 300~K of the armchair SLMoS$_{2}$ with PBCs, which are used to eliminate edge effects. It shows that a strain of 0.08 can reduce the thermal conductivity by nearly $40\%$. Similar strain effect has also been found on the thermal conductivity of the graphene.\cite{JiangJW2011negf,WeiN} This considerable reduction of the thermal conductivity is mainly due to the strain-induced scattering of the acoustic phonon modes. The application of the uniform axial mechanical strain effectively excites the low-frequency longitudinal acoustic phonon modes. These effective longitudinal acoustic phonons do not assist the thermal transport, but they participate in the scattering of other phonons. As a result, the thermal conductivity is considerably reduced.

\section{conclusion}
In conclusion, we have obtained a parameterization of the SW potential for the interatomic interactions in SLMoS$_{2}$.  The SW parameters are obtained by fitting to the phonon spectrum of the SLMoS$_{2}$. In particular, the SW potential is able to reproduce the energy gap around 250~{cm$^{-1}$} in the spectrum and the cross-over between the two highest-frequency intra-layer breathing phonon branches. It  also provides an accurate prediction for the Young's modulus of SLMoS$_{2}$. The SW potential was then used to predict the chirality and size dependence for the Young's modulus of the SLMoS$_{2}$ nanoribbons with free edges, where a strong reduction in Young's modulus with decreasing size was observed. Finally, we performed molecular dynamics simulations to study the thermal transport in SLMoS$_{2}$. We find that the free edges in the SLMoS$_{2}$ nanoribbon are not stable and can induce a melting phenomenon at temperatures far below the melting temperature of bulk MoS$_{2}$.  The temperature and strain dependence of the thermal conductivity was studied, and in particular a substantial decrease in the thermal conductivity with increasing tensile strain was observed.

\textbf{Acknowledgements} The work is supported by the German Research Foundation (DFG).  HSP acknowledges support from the Mechanical Engineering Department of Boston University.


%

\end{document}